\documentstyle[12pt,emlines]{article}

\newcommand{\f}{\phi}

\newcommand{\w}{\omega}

\newcommand{\be}{\begin{equation}}
\newcommand{\ee}{\end{equation}}
\newcommand{\mb}{\bf}

\title{Critical Behavior of the Water Coupled
 to a Local External Perturbation }

 \author{Dmitri Volchenkov and Ricardo Lima \\
{\small \em  CNRS, Centre de Physique Theorique, Luminy Case 907,} \\
{\small \em 13288  Marseille Cedex 09,  France  }\\
{ \small \em E-mail volchen@cpt.univ-mrs.fr, lima@cpt.univ-mrs.fr}}

\begin{document}
\thispagestyle{empty}
\noindent
 \maketitle

\begin{abstract}

The response  of inviscid incompressible
 unbounded fluid subject to a localized external
  perturbation is studied.  The physically relevant
 hypotheses on the mode coupling mechanisms is justified
by renormalization-group method.
 The scaling laws for the scalar and vector
velocity potentials are derived.
The spectrum of energy of perturbed fluid versus
the distance apart from perturbation is computed.

 \end{abstract}

PACS number(s): 03.40 G, 47.55, 47.10. +g, 05.40. +j, 42.27. Gs

\section{Introduction}

The excitation of wave motion in a fluid by various perturbative
 factors is a long-standing problem in hydrodynamics.  Particularly,
 the  wave generation on the water surface by turbulent
air flow  over the surface and the rising of tsunamy-like waves excited
 by bottom earthquakes are of current interest of modern ocean studies.
 The physical mechanisms coupling the perturbation and water
  must be understood in order to describe the wave motions on the
 ocean surface. In spite of great success in the description of
 statistical properties of water waves which have been achieved
recently in the framework of so-called Zakharov formalism
 (see, for example \cite{1} and  references therein) many phenomena
on the ocean surface   remain unclear.  The great advances in study of
surface phenomena were achieved in the modern critical phenomena theory
 recently \cite{2}.  They are due to employ of quantum field theory
 methods and quantum field renormalization group method (RG) \cite{3}
 in particular which are designed to  describe systems with an infinite
number of degrees of freedom.  The results indicate convincingly
 that the analogous approach would also be of great advantage in the
 explanation of ocean surface phenomena. We intend to propose the
statistical description of ocean surface  phenomena within the framework
of modern critical phenomena theory \cite{4}.  In the present paper we
 consider a simple physical model coupling water to perturbation
 which can be adopted for the use of RG method.  Our aim is to outline
 the basic properties of such a model for what purpose we now limit our
 consideration to the case of inviscid incompressible unbounded fluid.
 The work is concerned with the large-distance and long-time asymptotic
 behavior of water response for an external localized perturbation
risen by a pressure field pulse; it is imposed that the scale of
 perturbation $l_p$ is much less than the main scale of the problem
$r\gg l_p$.  The duration $\tau$ of energy input provided by the
 pressure pulse determines the scale of perturbed region
\be
 l_p=c\tau,
\label{tau}
 \ee
 where $c$ is a speed of perturbation spread in fluid.
  We are interested in the statistical properties of water
response in the range $r\gg l_p$ and $t\gg \tau$.
The crucial importance for the what following is that  the time
 derivative can be eliminated from hydrodynamic  equations,
 \be
 \mb {div}
{\ } \mb v(\mb x,t)=0,
 \quad \Delta p(\mb x,t )=-\partial_iv_j(\mb x,t)\partial_j v_i(\mb x,t),
  \label{eq}
  \ee
  where $ \mb v (\mb x,t)$ is
the velocity of fluid, $p(\mb x,t )$ is the field of pressure, and
 $\Delta$ is the Laplace differential operator.  We have taken in
(\ref{eq}) the constant fluid density to be $\rho_0=1.$ Summation
 over successive indices will be implied.
x Then from (\ref{eq}) one obtains an expression for the pressure
 $p(\mb x,t ):$
 \be
p(\mb x,t )=-\int_{V_p} d\mb y {\ } \frac {\partial_i
v_j(\mb y, t)\partial_j v_i (\mb y, t)}{|\mb x-\mb y|},
\label{p}
\ee
where the integration is brought about over the perturbed region
$V_p.$
It is essential that (\ref{eq}) is invariant with respect
 to an arbitrary time-dependent velocity shift:
  \be
  \mb v_a(\mb x,t)\to \mb v (\mb
  x+\mb s,t)-\mb a(t),
  \quad p(\mb x,t) \to p(\mb x+\mb s,t)
  \label{gi}
  \ee
  where $\mb a(t)$ is an arbitrary function of time decreasing at
  $t\to -\infty,$ and
  $\mb s(t)=$   $\int^t_{-\infty}{\mb a}(t'){\ }dt'.$
  This property expresses the Galilean invariancy of hydrodynamic
  equations.
The continuity equation in (\ref{eq}) shows that the velocity field
 $\mb v(\mb x,t)$ can be presented as a sum of two terms
 \be
 \mb v(\mb
x,t)= - \mb {grad} {\ } \f(\mb x,t)+ \mb {rot} {\ } {\mb A}(\mb x,t),
\label{vel}
\ee
where $\f(\mb x,t)$ and ${\mb A}(\mb x,t)$ are the
scalar and vector potentials consequently.  The
formulation of hydrodynamic  equations in favor
 of the potentials $\varphi=\{\f, A_i\}$
elucidates the invariancy of the equations with
respect to the shift of the vector potential $A_i,$
 \be
  A_i({\bf x})\longmapsto
A_i({\bf x})-\partial_i \Lambda({\bf x})
 \label{A}
 \ee
 in which $\Lambda({\bf x})$ is an arbitrary scalar function.  This
invariancy expresses the {\em gauge} symmetry of hydrodynamic equations.
  In the present paper we shall consider the consequences of this
invariancy for the hydrodynamic  equations (\ref{eq}) for the
 inviscid unbounded incompressible fluid.
Significant results on stationary spectra of the fully
 developed water response
for the perturbative pulse may be derived phenomenologically
in analogy with the fully developed turbulence theory \cite{Monin} and
the statistical theory of waves based on the Zakharov's kinetic
equations \cite{MK}.  This approach is related to some conserved
 quantities, {\em i.e.,} the wave action, energy, or momentum
and to the idea of localization of pumping and dissipation ranges in
 separated parts of scale spectrum.  By the way, a stationary spectrum
can have place for some scale interval which is transparent for the
 current of some quantity.
For example, one can point out the well-known {\em inertial}
range of
 Kolmogorov in the theory of fully developed turbulence which is
transparent for the energy current from the large-scale region of
 energy pumping to the small-scale range of viscous dissipation.
In the
Kolmogorov's inertial range the energy distribution versus
 wave-numbers
is described by the Five Thirds Kolmogorov's Law.
In case of the fully developed turbulence the Law of Five Thirds
 describes the only possible spectrum since the three-dimensional
isotropic
movements of the inviscid incompressible fluid preserve the only integral
quantity, {\em i.e.,} the energy.  However, in the problem
discussed there are variety of spectra to be realized, since in the process
of perturbation spread there are two more conserved integral
quantities:  {\em the net momentum} of fluid and {\em the enstrophy}.
 In case of an isotropic, $O(3)$-symmetrical, perturbation the net
momentum of fluid is equal to zero, however, even in this case there
are two different spectra which are determined by the energy ${\cal W}-$
 and the enstrophy ${\cal E}-$currents.  The velocity spectrum which is
 determined
 by the energy current from the pumping region $k_p=0$ into the
dissipation region $k_d=\infty$ is the well-known spectrum of the
 fully developed turbulence theory,
 \be
 {\cal W}^{1/3}k^{-1/3}\simeq v(k).
\label{v}
 \ee
 However, in the case considered $k_p>k_0\simeq 1/l_p \ne 0$
and $k_d\gg k_0$,
  so that the relevant inertial range lies apart
from the scale spectrum $k_0> k \simeq 1/r$, and (\ref{v})
 cannot have place in the problem considered.
 The enstrophy current (the squared averaged vorticity)
 determines the spectrum for the vector potential field
  $A$ in the form
   \be
   {\cal E}^{1/2}k\simeq A(k)
\label{as}
\ee
which transparency interval
is exactly the interval in question.
In language of the critical phenomena
theory the spectrum (\ref{as}) determines the {\em critical dimension }
 $\Delta[A ]$ of the field $A$, $\Delta[A ]=1$.  The use of critical
dimension allows to compute the spectrum of any correlation function of
 the field $A$ by simple dimensional counting.  For example, for
the pair correlation function in Fourier representation
 $D_A(k)\equiv \langle\mb A(\mb k)\mb A(-\mb k)\rangle,$ one
obtains an asymptotics:
   \be
D_A(k)\sim k^{\Delta[D_A ]},
 \quad \Delta[D_A ]=2\Delta[ A ]-d,
 \label{da}
 \ee
  where $d$ is the dimension of space.
For practical purposes, however, it is important to discuss
not the correlation function (\ref{da}) but a one-dimensional
spectrum
\be
\hat D_A=\frac
{S_d}{2(2\pi)^d}k^{d-1}D_A(k)
\label{1ds}
 \ee
 in which $S_d\equiv 2\pi^{d/2}/\Gamma(d/2)$ is the square
  of surface of the  unit sphere in
$d-$dimensional space.  For (\ref{1ds}) one obtains
 from (\ref{da}):
 \be
 \hat D_A \sim k.
 \label{tda}
 \ee
 Furthermore, the  phenomenological considerations analogous to (\ref{as})
 readily allows to determine  the  spectra of some quantities
 which can be measured in experiments,
 for example, for the energy as a function of  distance from
 the perturbation point,
  $E(r),$ we obtain
\be
{\cal E}r^{-4}\simeq E(r).
\label{es}
\ee
 From  phenomenology, however, it is not possible to fix
the spectra of  all quantities simultaneously. For example, one has
 nothing
to say about  the scalar potential spectrum $\f (k)$. To determine it
 one can
add some extra assumptions on the character of coupling
 mechanisms between
 different modes of fluid motions.  Actually, the formulation of
such a hypotheses is the crucial point of the problem of coupling
 water to perturbations.
Clearly, these additional assumptions will sufficiently depend on the
  geometry of perturbation. As we have seen above, in some cases
the perturbation pulse can rise a fluid current, then the net momentum
of fluid  $P$ will be nontrivial. The new  quantity conserved in
 the scale interval
of the problem will leads to another  possible spectrum.
In the present paper we discuss the case of $O(3)-$symmetrical perturbation;
  we   consider the case with a current aligned in the  fluid
in the forthcoming paper. Even for the simplest case of $P=0$
 the formulation of   physically relevant hypotheses on coupling water
 to perturbation
is a  nontrivial problem due to an infinite number of
 degrees of freedom.
 To solve it we apply the RG approach under the following reason:
 In the framework of RG method the physical degrees of freedom are to be
 replaced by the scaling degrees which are related to the physical degrees
 through the RG transformations of fields and parameters of the theory.
Since the properties of scaling degrees of freedom possess
  a group structure
 (the renormalization group) one investigates them much easier
 then those of
 the origin problem. The results obtained from RG-analysis
 are considered as
   somewhat statistical steady state limit of the physical system.
  The renormalized correlation functions are
distinguished from   their physical analogies only by normalization
conditions, so that they  can be also used for the analysis of
  asymptotic properties of the physical
system.
The plan of the paper follows: In the Section 2 we consider
 the direct consequences of
 the symmetry (\ref{A}) for the
 hydrodynamic  equations (\ref{eq}). As a result we derive
an effective action functional which   allows to understand
the water response on the local
 perturbation as a critical phenomenon. The action functional
 presented has the most general form
 and then needs to be supplied by an assumption on the coupling mechanism.
 To formulate the physically relevant hypotheses we develop the RG-analysis
  in the Sections  3, 4, and 5 consequently.
 We show that in the symmetrical case considered ($P=0$)
  the statistical properties
 of water response  are managed by the dynamics of
  vector potential $\mb A.$ This
 conclusion instantly fixes the spectra of all quantities of the theory.
 In the Sections 6 and 7 we discuss the various composite
operators of the theory
 which are responsible for the amendments to critical scaling and for
 the spectra of some quantities which can be measured experimentally.
 In particular,
 in the Section 7 we compute the one-dimensional energy spectrum $E(r)$
 as a function of distance from the point of perturbation.
 The results obtained for the spectra of the pair correlation function
 $D_A$ and  the energy $E$ meet the
 phenomenological relations (\ref{tda})  and (\ref{es}).
We believe that the obtained results would give a key for better
understanding
of the origin of a numerous ocean phenomena such as Lengmuir
circulations, spin-off eddies, and so on which theoretical
justification have not given by this time.

\section {The Description of Model and the Effective Action Functional}

In this section we formulate a real time classical field theory
  of equations (\ref{eq}) in the Lagrange formalism and a
model of nonlinear coupling mechanism between different wave modes.
 As a result we derive the effective action functional of
{\em abelian gauge-invariant} field theory which describes
 the statistical properties of long-range water response on
the external perturbation.  We
consider now the basic properties of the theory following from
 the symmetry (\ref{A}).

(i) First, the eddy
component of   velocity field is expressed
 by a {\em gauge} invariant tensor,
\be
 F^k=(\mb {rot } {\ }\mb
A)_k=\partial_i A_j-\partial_j A_i.
\label{F}
\ee

(ii) Second, in accordance to the Noether's theorem the symmetry
(\ref{A}) relates to a conserved current, {\em vorticity},
\be
\partial_i
J_i=0, \quad \partial_t J_i=0,
\label{cc}
\ee
 where $ J_i=\partial_jF^k-\partial_kF^j$, $ (i\ne k\ne j).$

(iii) Third, the classical equations (\ref{eq}) do not
lead to a hamiltonian in the usual way:  as a direct consequence of
 gauge
invariance, the equations (\ref{eq}) do not depend on time derivative.
  It is therefore impossible to define the conjugated momenta, \cite{4}.
Nevertheless, (\ref{eq}) are just the field equations which can
be derived from the classical lagrangian ${\cal L} (\varphi,p):$
\be {\cal L}
(\varphi, p)= \frac 12 \int d{\!}\mb x { \ } \left[ (\partial p)^2 +
 (\partial\f)^2 + \frac 12 F^2 + p\partial_iv_j\partial_jv_i + J_iA_i
\right],
 \label{L}
 \ee
 where we have introduced the tensor $v_iv_j\equiv
(\partial_i\f)(\partial_j\f)+F^iF^j- \partial_i\f F^j-\partial_j\f
F^i.$
The statistical properties of mechanical system can be
 derived from the partition function of statistical
  mechanics $Z=Tr(e^{-S})$ with
somewhat classical dimensionless action $S$.
In case of an infinite number of degrees of freedom
 one can write down the partition function in
functional integral representation:
 \be
  Z(J_i)= \int [d \varphi][dp]
  \exp\left[ -S (\varphi, p)+\int d{\!}\mb x {\ }dt J_i({\bf
x},t)A_i({\bf x},t)\right],
\label{Z}
\ee
in which the euclidean static action $S(A_i,\f)$ has the form
 \be
  S (\varphi, p)=\frac 12 \int
d{\!}\mb x {\ } \left[ (\partial p)^2
+ (\partial\f)^2 +
 \frac 12 F^2_{ij}
  + p\partial_iv_j\partial_jv_i\right].
  \label{s}
  \ee
   Here we note
that the quadratic part of the action functional (\ref{s})
 which is relevant to a free theory, {\em i.e.,} with no
 coupling between water
and perturbation is not symmetrical with respect to Galilean
transformation (\ref{gi}) but is symmetrical with respect to the following
transformations
 \be
\left\{
\begin{array}{l} A_i({\bf x})
 \longmapsto A_i({\bf x})-\partial_i \Lambda({\bf x}), \\
  \f({\bf x})
\longmapsto \f({\bf x})e^{iu_0\Lambda({\bf x})},
 \end{array}
  \right.
   \label{A6}
    \ee
where $\Lambda({\bf x})$ as usual is an arbitrary scalar
function, and $u_0$ is a potential coupling constant.
  These transformations expresses the so-called {\em $U(1)-$gauge
symmetry} ($U(1)$ is the group of multiplication by complex numbers).
 The last term in (\ref{s}) does not meet the entire symmetry
(\ref{A6}) but preserves the simple gauge symmetry (\ref{A}).
This symmetry breaking shows that the pressure field as it is included in
 the
action functional contains somewhat redundant degrees of freedom.

 Instantly close to the region of perturbation the pressure pulse
  rises the wave motions with eigenmodes $k> k_0\simeq 1/l_p.$
   Due to strong
nonlinearity of the interaction in the Navier-Stockes equation
 the eigenmodes of oscillations spread very fast from a band of order
$l_p^{-1}$ over the whole spectrum, and various multipole oscillations
of any type are arisen with time.  One can say that after a short
period of time motions of any modes are present in the water.
 Clearly, the long-distance fluid behavior
 will depend to some extent on the
statistical properties of   wave mode coupling.

Assuming an effective action functional to be

(i) {\em local} in space and time, {\em i.e.,}
it depends only
 on the fields ${\bf A} (x)$ and $\f (x)$ and their
partial derivatives
(and not on products of fields and their
 derivatives at different points),

(ii) {\em invariant} under space and time translations, {\em i.e.,}
space and time coordinates do not appear explicitly in the action,
we now suppose the simplest model for the coupling mechanism by
  inclusion of the $\varphi^4$-type interaction term into
(\ref{s}):
\be
S (\varphi, p)=\frac 12 \int d \mb x {\ }
\left
[ (\partial p)^2+ (
\partial\f)^2 + \frac 12 F^2_{ij} +
p\partial_iv_j\partial_jv_i +\frac 13 g\varphi^4\right],
\label{f4}
\ee
 with a wave modes coupling constant $g$.
 In accordance with the
general critical phenomena approach we note
 that the accounting of
  highest oscillation harmonics, {\em i.e.,} $\varphi^6$, $\varphi^8$
and so on cannot alter the large-distance asymptotic behavior
 of water response for localized pressure pulse if $g\ne 0$ \cite{8}.

Since the pressure pulse is localized in the scale $ l<l_p,$
and the locality postulated for the action functional requires
redundant degrees of freedom for the fluctuating pressure
field $p(\mb x,t)$,
we therefore can integrate it over in the
 partition function (\ref{Z}).
The result of functional integration does
 not depend on  $p(\mb x,t).$
 Due to (\ref{p}) this procedure is of perfect clarity and is reduced
technically to elimination of the quadratic term proportional
 to $(\partial p)^2$ from (\ref{f4}) and to replacement of the
  pressure field
$p({\bf x},t)$ in the $U(1)-$ breaking term
$p\partial_iv_j\partial_jv_i$ by (\ref{p}).  In particular, it leads
 to a new term in
(\ref{f4}) of the form
 \be
 \frac 12
 \int d\mb x {\ }\partial_iv_j({\bf x},t)\partial_jv_i({\bf x},t)
 \int_{V_p} d\mb y {\ }
\frac{\partial_i v_j(\mb y, t)\partial_j v_i (\mb y, t)}{|\mb x-\mb y|},
 \label{t}
  \ee
   which relates the fluctuations of velocity fields risen by
the perturbation pulse in $V_p$ to the fluctuations apart from the
 perturbed region.  We note that (\ref{t}) preserves the entire
$U(1)-$gauge symmetry (\ref{A6}).

Performing the integration over $V_p$ and moving the derivatives
 onto the result of integration, one can rearrange (\ref{t}) to the form
  \be
\frac 12 \int d\mb x K_{ij}(\mb x)Q_{ij}(\varphi,\mb x, t),
\label{K}
 \ee
 where $Q_{ij}$ is a quadratic form of potentials $\f$ and $A_i,$
and $K_{ij}$ is a source kernel.  To explain the meaning of (\ref{K})
 we adduce some arguments from renormalization group analysis which we
are going to apply to the
 theory discussed in the forthcoming sections.
  For the future purposes of   RG analysis we need only insertions
at zero momentum and we could, in principle, restrict ourselves to
 constant sources $K_{ij},$ since the renormalized action which we
 derive for the case in question is obtained by setting $K_{ij}(\mb x)$
 as constant, \cite{4}.

Another important feature of (\ref{K}) stems from the fact that
 the RG transformations generate all possible linearly independent
  quadratic
terms in $Q_{ij}$,
{\em i.e.,} $\f^2(\mb x,t),$ $\f A_{i}(\mb x,t),$
and $A_iA_j(\mb x,t ).$ However,  some certain linear combinations of
quadratic operators in $Q_{ij}$ only  are relevant to scaling degrees
 of freedom and possess, by the way, the definite physical meaning.  We
compute all such linear combinations in the Section 6. Nevertheless,
 one can see before calculations that the   linear
combinations of quadratic operators which are of importance
 for the RG analysis
  have to be  $U(1)-$gauge symmetrical.
    There is only one such a combination,
     {\em i.e., }
      \be
      m^2\f^2(\mb x)
      \label{mf2}
      \ee
      in
which $m^2$ is somewhat mass parameter (the coefficient of
the relevant RG-invariant operator).
  The use of Ward identities which express the $U(1)-$gauge
   invariancy of the theory allows to show
that all other combinations of quadratic operators are
ultra-violet (UV) finite, {\em i.e.,} the relevant correlation
functions   do not have UV divergences.  Thus, with no loss of
 generality we can omit all   combinations except for (\ref{mf2})
 from the
consideration.  It leads to the action functional which follows
 \be
  S (\varphi )=\frac 12 \int d \mb x {\ }
 \left[ (\partial\f)^2 + \frac 12
F^2_{ij} + m^2\f^2 +\frac 13 g\varphi^4\right].
 \label{a}
 \ee
 The effective action functional (\ref{a}) which
 is arranged to describe the
asymptotic properties of water response   is not renormalizable,
 and it has
no solutions in
the massless limit, when $m^2=0$,\cite{4}.
  Again, the reason is
 that
it still has some redundant degrees of freedom, the gauge
degrees, with unknown dynamics.

To construct a renormalizable theory we are led to introduce
an abelian gauge geometrical structure.  By the way,

(i) $\f({\bf x})$ and $\f^*({\bf x})$ are vectors
for $U(1)$ transformations,

(ii) the derivative $\partial_i$ is replaced by the
 covariant derivative $\nabla_i$:
   \be
   \nabla_i=\partial_i+iu_0A_i,
   \label{A7}
    \ee
    where
$u_0$ is the coupling constant of interaction between
the scalar and rotational components of the
 velocity potential $\varphi$
 (analogous to
the electron charge $e$ in electrodynamics).

(iii) It follows that the curvature tensor
is $iu_0F_{ij}:$
 $$ iu_0F_{ij}=[\nabla_i,\nabla_j]=iu_0(\partial_iA_j-\partial_jA_i).$$

(iv) Since the $U(1)$-gauge group is abelian
 ($\bf A(\mb x,t)$ is a translation invariant), one can write
  the parallel transporter $U(C)$ along
any continuous contour $C$  which is an element of $U(1).$ In
 terms of a line integral:
   \be
    U(C)=\exp\left[-iu_0\oint_C A_i(s) {\ } ds_i\right]
\label{pt}
 \ee
 as a consequence of   vorticity conservation.
  Thus, the rotational component of   velocity potential just
   carries on the
fluctuations of the scalar potential field $\f({\bf x})$.
  By the way, two solutions for different points $\f({\bf x},t )$
  and $\f({\bf y},t)$ are related through the parallel
 transporter (\ref{pt}),   where $C$ is an integration path
 connecting the points
$\mb x$ and $\mb y$, \cite{4}.

The form of the action functional (\ref{a}) which
meets the geometrical structure (i)-(iv) follows
 \be
 S (\varphi )=\frac 12 \int d \mb x {\ }
 \left[ (\partial p)^2 |\nabla_i\f|^2 + \frac 12 F^2_{ij}
 + m^2\f^2 +\frac 13 g\f^4\right].
  \label{ef}
  \ee
  The problem of the gauge
invariant theory is that the local gauge invariant
action does not provide a dynamics to the gauge degrees
of freedom.  We take them into
account, performing the standard procedure analogous to
the Fadeev-Popov quantization \cite{4}.
 Let us write the gauge field $ A_i$ in
terms of a gauge field $B_i$ projection of $A_i$ on
 some gauge section,
{\em i.e.,} satisfying some gauge condition, and a gauge
transformation:
\be
 A_i=B_i+\partial_i \Lambda.
\label{gt}
 \ee
  We assume that this decomposition is unique.
   For $\Lambda ({\bf x})$
one imposes:
\be
 \partial^2 \Lambda ({\bf x})+\partial_iB_i({\bf x})=
h({\bf x}),
 \label{A8}
   \ee
   in which $h({\bf x})$ is a
stochastic field for which a probability distribution is given.
 We do not include a term proportional to $\f({\bf x})
e^{i\Lambda({\bf x})}$ in the condition (\ref{A8})
omitting the $U(1)$ gauge degrees of freedom to simplify
 the model and to avoid
the appearance of Fadeev-Popov ghost fields.
  Including the
equation (\ref{A8}) in the functional integral for the partition
 function, one
can see that since the result does not depend on the dynamics of
$\Lambda({\bf x})$ and on the field
 $h({\bf x})$ either, one can
integrate over $h({\bf x})$ with the gaussian measure:
\be
 [d\rho(h)]=[dh] \exp\left[ -\frac 1{2\zeta}
\int d{\!}x h^2({\bf x})
 \right]
  \label{mh}
\ee
in which $\zeta$ is an arbitrary valued ($ \zeta\in [0,\infty) $)
 auxiliary gauge parameter of the theory.

The resulting effective action of the model has the form:
  \be
  S (A_i,\f)=\frac 12 \int d{\!}x \left[ |\nabla_i\f|^2+
\frac 12 F^2_{ij}
+\zeta^{-1} (\partial_i A_i)^2+m^2\f^2+
\frac 13 g \f^4\right].
  \label{A9}
  \ee
In the forthcoming sections we show that the model
(\ref{A9}) demonstrates the existence of a statistically
steady state independently of the details of velocity evolution.
 The functional
(\ref{A9}) is analogous to a $U(1)-$invariant action for a
charged scalar field with a $|\f|^4$ self-interaction
(so called {\em Abelian
Higgs Model}, \cite{6}).  This theory allows the multiplicative
renormalization, \cite{7} and for some values of the parameters
$\{m_0,g_0,u_0\}$
the corresponding physical
 system tends to a steady state
in a large-distance limit and demonstrates a universal behavior.
 In language of
the critical phenomena theory, there is a nontrivial IR-stable
fixed point which determines the critical asymptotics of
correlation functions
of the fields $\f ({\bf x})$ and ${\bf A}({\bf x})$.
 The relevant critical index for $A$ meets the phenomenological result
(\ref{as}).

We conclude this section by an explanation of   physical
meaning of solutions for different signatures of the mass parameter $m^2.$
In case of ${\bf A}=0$ in (\ref{A9}), one has the standard model
of a scalar unharmonic oscillator.  This model is, may be, the most
popular and well-investigated action of the modern theoretical physics.
 For $m=0$ the oscillator is subject to a phase transition.
At the classical level in the symmetrical case ($m^2>0$) the oscillator
 model describes the fluctuations having the trivial expectation value
of the field, $\langle\f \rangle_0=0$ (see Fig. 1.a).
 If $m^2<0$, the system allows
two possible expectation values for the field $\langle\f \rangle_0=\pm
\sqrt{ m^2 /g}$ (see Fig. 1.b).
 The latter situation is usually referred to as spontaneously
 broken symmetry.

The physical consequences for the model considered can be
readily understood.  The signature of integral operator kernel
 $K_{ij}(\mb x,t)$
depends on the certain physical conditions and on the geometry
 of perturbation, and it is determined directly
  by the signature of operator
$\partial_iv_j\partial_jv_i.$ This signature does not depend
 neither on the certain value of velocity nor its evolution, but
  it depends on the
topological properties of fluid flow risen by the perturbation.

The case of symmetrical perturbation, when the net momentum of
fluid is equal to zero (for example, on the surface of a large
 scale eddy, see Fig.2. a) can
 be described by the model (\ref{A9}) with $m^2>0.$
 In the vicinity of saddle points,
 $\partial_iv_j\partial_jv_i<0,$ {\em i.e.,}
 when there is a net fluid current from the region of initial
perturbation
  into   outside (see Fig.2.b), and the net fluid
momentum $P\ne 0,$ one can use (\ref{A9}) with $m^2<0.$

In the what following sections of the present paper we shall consider
 the case of symmetrical perturbation, $m^2>0$, (Fig.2. a).
 The presence of the gauge field ${\bf A} $ affects sufficiently
 the behavior of the system due to the transporting role of the field
${\bf A}$.  We demonstrate below that the ordinary infrared
 (IR)-stable fixed point of  RG transformations which is responsible
  for   self-similar behavior in the standard $\f^4$-theory turns
  out to be unstable, however, the new fixed point acquires stability
 for the real value  $\epsilon_r=1/2$ of the parameter
   $2\epsilon = 4-d.$ We shall derive   formulae
for universal scaling profiles to the first order in $ \epsilon .$
The profiles has a power-law behavior for the
large distances $r\gg l_p.$

\section{Infrared Singularities of Perturbation Theory Diagrams}

In the present Section we develop the diagram technique relevant
 to the theory (\ref{A9}) and discuss the large-distance
 ($k\to 0,$ in  momentum representation) singularities of
 perturbation theory diagrams.

The model (\ref{A9}) can be considered in the $d$-dimensional
space $\mb x$ with UV-cut off $\Lambda\equiv k_0\simeq 1/l_p$.
 Each quantity
in (\ref{A9}) corresponds to one (momentum) canonical
 dimensionality $d_f$, which is completely determined by
 the space dimensionality $d$.

In the critical phenomena theory one seeks the asymptotic
for  correlation functions in the region $k$, $m\ll \Lambda$ for
 which one considers
$g_0\simeq u_0^2\simeq const {\ } \Lambda^{4-d}$ with $const \leq 1.$
 From now on we supply all the parameters in (\ref{A9}) by the lower
index $"0"$ to distinguish them from those in renormalized action
 forthcoming.
  In order to be specific we consider the pair correlation
functions of the potentials
$D_\f(r)= \langle\f(\mb x)\f(\mb y)\rangle
$ and $D_A(r)=
 \langle A(\mb x)A(\mb y)\rangle$ in which $r=|\mb
 x-\mb y|$.

In  momentum representation these correlation functions are found from the
Dyson equations ($p$ is the external momentum)
\be
D^{-1}_\f=p^2+m_0^2-\Sigma_\f(p),
\quad D^{-1}_A=p^2+i\varepsilon-\Sigma_A(p),
\label{27}
 \ee
 where $\Sigma_\f(p)$ and $\Sigma_A(p)$ are the
infinite sums of all 1-irreducible Feynman diagrams
(see Fig.3) whose vertices correspond
 to the multipliers $g_0$ and $u_0^2$

\vspace{1cm}

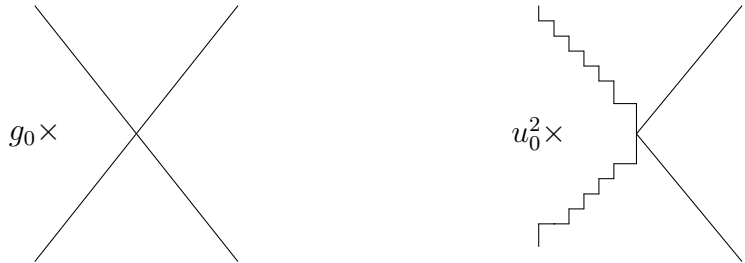
\unitlength=1mm
\special{em:linewidth 0.4pt}
\linethickness{0.4pt}
\begin{picture}(125.00,39.00)
\emline{45.00}{4.00}{1}{45.00}{4.00}{2}
\emline{31.00}{39.00}{3}{58.00}{5.00}{4}
\emline{58.00}{39.00}{5}{31.00}{5.00}{6}
\put(31.00,22.00){\makebox(0,0)[cc]{$g_0\times$}}
\emline{125.00}{39.00}{7}{111.00}{22.00}{8}
\emline{111.00}{22.00}{9}{125.00}{5.00}{10}
\emline{111.00}{22.00}{11}{111.00}{26.00}{12}
\emline{111.00}{26.00}{13}{108.00}{26.00}{14}
\emline{108.00}{26.00}{15}{108.00}{29.00}{16}
\emline{108.00}{29.00}{17}{106.00}{29.00}{18}
\emline{106.00}{29.00}{19}{106.00}{31.00}{20}
\emline{106.00}{31.00}{21}{104.00}{31.00}{22}
\emline{104.00}{31.00}{23}{104.00}{33.00}{24}
\emline{104.00}{33.00}{25}{102.00}{33.00}{26}
\emline{102.00}{33.00}{27}{102.00}{35.00}{28}
\emline{102.00}{35.00}{29}{100.00}{35.00}{30}
\emline{100.00}{35.00}{31}{100.00}{37.00}{32}
\emline{100.00}{37.00}{33}{98.00}{37.00}{34}
\emline{98.00}{37.00}{35}{98.00}{39.00}{36}
\emline{111.00}{22.00}{37}{111.00}{18.00}{38}
\emline{111.00}{18.00}{39}{108.00}{18.00}{40}
\emline{108.00}{18.00}{41}{108.00}{16.00}{42}
\emline{108.00}{16.00}{43}{106.00}{16.00}{44}
\emline{106.00}{16.00}{45}{106.00}{14.00}{46}
\emline{106.00}{14.00}{47}{104.00}{14.00}{48}
\emline{104.00}{14.00}{49}{104.00}{12.00}{50}
\emline{104.00}{12.00}{51}{102.00}{12.00}{52}
\emline{102.00}{12.00}{53}{102.00}{10.00}{54}
\emline{102.00}{10.00}{55}{100.00}{10.00}{56}
\emline{100.00}{10.00}{57}{100.00}{10.00}{58}
\emline{100.00}{10.00}{59}{98.00}{10.00}{60}
\emline{98.00}{10.00}{61}{98.00}{7.00}{62}
\put(98.00,22.00){\makebox(0,0)[cc]{$u^2_0\times$}}
\end{picture}

and whose lines correspond to the bare propagators

\vspace{1cm}

\unitlength=1mm
\special{em:linewidth 0.4pt}
\linethickness{0.4pt}
\begin{picture}(86.00,52.00)
\emline{20.00}{41.00}{1}{23.00}{46.00}{2}
\emline{23.00}{46.00}{3}{25.00}{41.00}{4}
\emline{25.00}{41.00}{5}{28.00}{46.00}{6}
\emline{28.00}{46.00}{7}{31.00}{41.00}{8}
\emline{31.00}{41.00}{9}{34.00}{46.00}{10}
\emline{34.00}{46.00}{11}{37.00}{41.00}{12}
\emline{37.00}{41.00}{13}{40.00}{46.00}{14}
\emline{40.00}{46.00}{15}{43.00}{41.00}{16}
\emline{43.00}{41.00}{17}{46.00}{46.00}{18}
\emline{46.00}{46.00}{19}{49.00}{41.00}{20}
\emline{49.00}{41.00}{21}{52.00}{46.00}{22}
\emline{52.00}{46.00}{23}{55.00}{41.00}{24}
\emline{55.00}{41.00}{25}{58.00}{46.00}{26}
\emline{58.00}{46.00}{27}{61.00}{41.00}{28}
\emline{61.00}{41.00}{29}{64.00}{46.00}{30}
\put(21.00,52.00){\makebox(0,0)[cc]{$A_i(\bf k)$}}
\put(59.00,52.00){\makebox(0,0)[cc]{$A_i(-\bf k)$}}
\put(71.00,43.00){\makebox(0,0)[cc]{$\equiv$}}
\put(91.00,43.00){\makebox(0,0)[cc]{$P_{ij}(\zeta,k)(k^2+i\varepsilon)^{-1},$}}
\emline{20.00}{10.00}{31}{64.00}{10.00}{32}
\put(20.00,17.00){\makebox(0,0)[cc]{$\phi(\bf k)$}}
\put(59.00,17.00){\makebox(0,0)[cc]{$\phi(-\bf k)$}}
\put(71.00,11.00){\makebox(0,0)[cc]{$\equiv$}}
\put(86.00,11.00){\makebox(0,0)[cc]{$(k^2+m^2_0)^{-1},$}}
\end{picture}

\vspace{1cm}

where $P_{ij}(\zeta,k)\delta_{ij}+(\zeta-1)k_ik_j/k^2$ is
the gauge dependent projector, and $\varepsilon$ is a
regularization parameter
for $k=0.$ We investigate the theory in $d=4-2\epsilon$ dimensions
  considering $\epsilon$ as a small parameter of a regular expansion
   which has
 $2\epsilon_r=1$ as an actual value.  For $0<2\epsilon<1$ the
  diagrams in Fig.3, independent of $p$ and $m_0$, have the
  algebraic
 UV-divergent terms $\sim \Lambda^{2-2n\epsilon}$
(where $n$ is the order of perturbation theory)   corresponding
  to the simple shift of
 $m_0^2$ which does not alter the signature of $m_0^2$.
   If we consider the value of $m_0^2$ to be known exactly,
    it is necessary
  to discard all such terms.  As usual \cite{3},
    this is
 implemented by subtracting their values for $p=m_0=0$ from
 all graphs of Fig.3.  After these subtractions the integrals
  for $0<2\epsilon<1$
 become UV-convergent, the cut off $\Lambda$ can be eliminated
 (taken as $\infty$, {\em i.e.,} the initial perturbation
 scale $l_p$ is taken
 as zero), and the series Fig.3 takes
  the form
  \be
  D^{-1}=(p^2+m_0^2)\left[1+\sum_{n,l=1}^{\infty}
(g_0^nu_0^{2l}p^{-2\epsilon})^{n+l}
 c_{n,l}(m_0/p,\epsilon)\right].
 \label{28}
 \ee
 For $p\sim m_0\ll \Lambda$ and $\epsilon>0$ the dimensionless
  parameter of the expansion
 $g_0p^{-2\epsilon}\sim (\Lambda p^{-1})^{2\epsilon}$ in (\ref{28})
  is not small, and it is necessary to sum  the series.  This problem is
 solved by the RG method.

In a clearer formulation this problem reduces to a determination
of the asymptotic value of the propagator $D_{\chi}=D(\chi p,\chi m_0)$
 for
$\chi\to 0$ (everything is fixed except for $\chi$).
 This procedure is nontrivial for $\epsilon>0$ due to presence
  in the $c_{n,l}$ of poles
in $\epsilon$ and leads to the equations of RG which we shall consider
in the next Section.

Another problem, which is occurred in the region $m_0\ll p,$ is
connected with singularities of the coefficient $c_{n,l}$ in
 (\ref{28}) for $m_0/p\to 0$ and cannot be handled by RG.
This problem originates from
the finiteness of the physical value of $\epsilon$.  After
 removing of UV
divergences from diagrams of Fig.3 there are still diagrams
diverging for $m\to 0$ for any $\epsilon>0.$ This problem
had been discussed in
the critical phenomena theory where the method of short distance
 expansion (SDE) was employed, \cite{3}-\cite{4}.
  We shall apply SDE  to compute the leading amendments to
  critical scaling  of (\ref{A9}) with $m^2>0$ in the Section 6.

Finally, we make a note on the particular features
of perturbation series Fig.3 for the gauge invariant theory.
 Some diagrams in Fig.3 have
$\zeta-$dependent poles which are unphysical,
 since they have been
 introduced to make the theory renormalizable.
 The renormalization
constants of the gauge invariant theory as we
 shall define them later
 on are gauge independent, therefore we can fix the value of the gauge
parameter $\zeta$ in certain calculations.  In particular, we use
the Landau gauge ($\zeta=0$), so that the
 gauge field propagator is simply
proportional to the transversal projector $P_{ij}.$

\section{Renormalization-Group
  Equations.  Scaling Degrees of Freedom}

Now we discuss the renormalization procedure for
the model (\ref{A9}) and produce the renormalized action
functional, then we derive the RG
equations for renormalized correlation functions.  Renormalizability
of the theory (\ref{A9}) (the Abelian Higgs Model) for any value of the
gauge parameter $\zeta<\infty,$ is proven (see for example \cite{7}), and
we do not discuss it in details.
The UV-divergences (in our case the poles in $\epsilon$ in diagrams)
of the model considered are removed by the multiplicative
renormalization procedure.  It amounts to the following:
the initial
action  is referred to as nonrenormalized, its parameters and
coupling constants are referred to as bare; these are considered
 as some functions (remaining to be determined) of new renormalized
parameters and coupling constants.
The renormalized action functional
\be
\begin{array}{c} S_R(\mu,g,u,\zeta) =
\frac 12 \int dx
\left[Z_1((\partial\f)^2+\mu^{2\epsilon}u^2\f^2A^2)+
\frac 12 Z_2F^2+\right.\\
 + \left. \zeta^{-1}(\partial A)^2+ Z_3 m^2\f^2+\frac 13 Z_4
\mu^{2\epsilon} g \f^4\right],
 \end{array}
 \label{29}
 \ee
 is a function of renormalized coupling constants and parameters:
   \be
\begin{array}{c}
 g_0=\mu^{2\epsilon}gZ_g,\quad u_0^2=\mu^{2\epsilon}u^2Z_u,
 \quad m_0^2=m^2Z_m,\\
  \f^2=Z_\f \f^2_R,\quad A^2=Z_AA^2_R,\quad
\zeta_0=Z_\zeta\zeta_R,
\end{array}
\label{30}
\ee
where all renormalization constants $Z_a$ are the functions
 of four independent
quantities $Z_{1-4}:$
 \be
  Z_1=Z_\f,
\quad
Z_u=Z_\zeta=Z_A^{-1}=Z_2^{-1},
\quad Z_mZ_\f=Z_3,\quad Z_4=Z_gZ_\f^2,
 \label{31}
  \ee
  which can be
calculated within the framework of diagram technique.  We chose
the simplest form of subtraction scheme, where the divergences are
 presented as the bare poles in $\epsilon$ (so called "minimal
 subtraction
scheme"); $\mu\simeq1/l_p$ is the renormalization mass parameter,
 $g,\zeta, m$ and $u$ are renormalized analogies of the bare parameters
$g_0,\zeta_0, m_0$ and $u_0,$ $Z_a=Z_a(g,\epsilon,u,d )$ are
the renormalization constants.
  Due to gauge invariance of the theory the
terms  breaking the gauge symmetry are not renormalized, and they
 do not require counterterms, \cite{4}. The renormalized correlation
 functions $W_R$ meet the relation
 \be
 W^R(g,u,\mu)Z^{N_\f}_\f Z^{N_A}_A=W(g_0,u_0)
 \label{32}
 \ee
 in which $W^R$
are UV-finite functions
(they are finite in the limits $\epsilon\to 0$) for fixed
 parameters $a$.

The RG equations are written for the functions
$W^R$ which differ from the initial $W$
only by normalization and then can be used equally
validly for critical scaling analysis.
  To derive these equations  one  notes that
the requirement of eliminating singularities does not
determine the functions $e_0=e_0(e,\epsilon),
\quad e=\{g,u,m\},$ uniquely because of
 the value of $\mu$ is not fixed by any physical
condition.
Variation of $\mu$ for fixed values of bare
parameters $e_0$ leads to variations
 of $e$ and renormalization constants
(\ref{31}).  Following the standard notation,
 we denote by $D_\mu$ the differential operator
 $\mu D_\mu$ for fixed $e_0$.  Applying it on
both sides of (32) leads to the basic RG equation, \cite{5}:
  \be
\left[D_\mu+\beta_g\partial_g+\beta_u\partial_u-
\gamma_mD_{m^2}\right]W^R=0,
 \label{33}
 \ee
 where  we have used $D_x\equiv x\partial_x$
for any parameters of the renormalized theory;
for any $Z_i$
 \be
 \gamma_i\equiv D_\mu \ln Z_i,\quad \beta_\alpha \equiv D_\mu
\alpha,\quad \alpha \equiv \{g,u,\},
\quad i\equiv \{g, u, \zeta, m, \f, A\}.
 \label{34}
 \ee
These identities determine the $\beta$-functions of the
 theory considered,
\be
 \beta_g=-g\left[2\epsilon+\gamma_g\right],
\quad \beta_{u }=-u^2\left[2\epsilon+\frac 12 \gamma_{u}\right]
 \label{35}
 \ee
and the anomalous
dimensionalities $\gamma_i.$
One calculates the renormalization constants $Z_{1-4}$
 from the diagrams of perturbation theory (these
calculations are completely analogous to the relevant
 computations in  $f^4$-theory of the critical
 phenomena theory, \cite{4}) and then, using (\ref{34}),
 $\gamma_i-$ and
$\beta_\alpha$-functions. By the way,
all $\gamma-$ and $\beta-$ functions are constructed
 as series in $g$ and $u,$ and the functions $\gamma_i$ do
not depend on $\epsilon.$
Furthermore, the  relations between renormalization
 constants (\ref{31}) lead to
analogous relations for $\gamma_\alpha$:
\be
 \gamma_g=\gamma_4-2\gamma_1, \quad
 \gamma_u=-\gamma_2.
\label{ggu}
\ee
We have computed the relevant renormalization
constants up to the second order diagrams of perturbation theory.
These  computations are pretty standard, so that we just bring about
the results for $\gamma_i$ for the three dimensions,
\be
\gamma_1=\frac 16 g'^2+u'^2,\quad \gamma_2=6u'^2,\quad
\gamma_3=g'^2+6u'^2,\quad
\gamma_4= 6g'^2+\frac 73 g'+4u',
\label{gamma}
\ee
where $g'=g/16\pi^2$ and  $u'=u^2/16\pi^2.$ From (\ref{gamma})
one obtains the explicit expressions for the $\gamma$-functions of
fields and the mass:
 \be
 \gamma_\f=\frac 16 g^2+u'^2, \quad \gamma_A=6 u'^2,\quad
\gamma_m=-g'-\frac 23 u'+ \frac 56 g'^2+5u'^2.
  \label{37}
 \ee
Substituting (\ref{gamma}) into (\ref{ggu}) and (\ref{35}),
we obtains the expressions for the $\beta-$ functions:
  \be
\beta_{u }=-u'\left[2\epsilon-6u'^2\right],
\quad \beta_g=-g'\left[2\epsilon +
\frac 73 g'+4u'+\frac{17}{3}g'^2-2u'^2 \right].
 \label{36}
 \ee
Eight fixed points of the  RG transformation are
 determined by the system of equations
$\beta_\alpha(g'^*,u'^*)=0 $.  A fixed point is  stable with respect to
large-distance asymptotics if  the matrix $\w_{ij}\equiv
\partial_i\beta_j$ is a positively defined
matrix at the fixed point.
Fixed points and their stability regions with respect to the
 large-distance asymptotics are collected in the Tab. 1.

The fixed point $N$ 4 corresponds to the asymptotic behavior
of the scalar
 $\f^4$ model of  critical phenomena, \cite{8}.  In the model
discussed it is unstable due to scalar potential coupling to the vector
 potential field $\mb A$.  For the "physical" value of the parameter
$2\epsilon=4-d$ the point $N$ 3 is the only stable fixed point.  The fixed
 point $N$ 7 would be stable close to the four dimensions also.
The large-distance asymptotic behavior of  water response in
 three dimensions
 is then governed by the fixed point $N$ 3.  The inequality
$0<\epsilon<3$ determines the relevant basin of attraction
 in space dimensionality.

In the framework of  RG-approach the physical degrees of freedom
are replaced by the scaling degrees including anomalies.
 In particular,
the scaling functions are obtained in the form of the power
 series in $g$
 and $u$.  Asymptotically, these coupling constants are replaced by
their values in fixed points of RG-transformation.
The critical indices
 in the $4-2\epsilon $ expansion are obtained from the
$\gamma-$functions (\ref{37}) with  replacement of $g$ and $u$
 by $g^*$ and $u^*$ also.  By the way, the
properties of scaling degrees of freedom (see Tab.1)
 yield  some qualitative conclusions on the physical
properties of the  model.

From the data of Tab.1 it follows that for $N$ 3 the
 scaling degrees of freedom related to the $\f^4-$coupling
 vanish ($g^*=0$) in three
dimensions.  Thus, all diagrams proportional to $g$ are vanished
in  the symmetrical phase of (\ref{A9}), however, no one correlation
 function
becomes trivial. Considering the theory in three dimensions,
 one can
eliminate the scalar wave mode coupling
 term $g\f^4$ from the action (\ref{A9}),
 since in case of $m^2>0$ it does not contribute
 to the large-distance asymptotics of
water response.
  One can say that three dimensional dynamics of an inviscid
incompressible fluid
 involved in eddy motion has somewhat short of physical
degrees of freedom to allow the coupling between the
 different scalar wave modes.
  $N$ 7 in Tab.1 gives us an evidence that the both
coupling mechanism introduced in (\ref{A9}) are of equal
importance in four dimensions.

Choosing the certain fixed point
of RG-transformation ({\em i.e.}, $N$ 3), we
 neglect  all  couplings
 between various wave modes in  benefit for
 the certain one which
is responsible for vorticity conservation (\ref{pt})
in the large-distance limit.
This conclusion expresses exactly that  additional assertion
which we have needed  to complete the phenomenological description of
the problem of coupling water to the  $O(3)-$symmetrical perturbation.

\section{Solution of RG-Equations.  Critical Scaling and
 Asymptotics for Pair Correlation Functions}

In the present Section we derive the solutions of RG
differential equations
 (\ref{33}) for $D_\f$ and $D_A$.  The use of standard
dimensional counting supplied by (\ref{31}) and  (\ref{37}) leads to the
 following expressions for the one-dimensional spectra of the theory:
\be
 D_\f(s)\simeq_{s\to 0}s^{2-d+\eta_\f}f_\f(s,g,u^2,z),
\quad D_A(s)\simeq_{s\to 0}s^{2-d+\eta_A}f_A(s,g,u^2,z).
\label{38}
\ee
The functions $f_\f$ and $f_A$ of dimensionless arguments:
$s=kl_p,$ $z\equiv m^2r^{1/\nu},$ $\eta_\f=d-2+\gamma_\f$ $=d-2+\epsilon/3,$
$\eta_A=d-2 +\gamma_A$ $=d-2+2\epsilon,$ and
  $1/\nu=2+\gamma_m=2-2\epsilon$
 meet the RG-equations of the type (\ref{33}) which allow
to find out their scaling  asymptotics ($s\to 0$).

Here we note that for $\epsilon_r=1/2$ the asymptotics (\ref{38})
 for $D_A$ meets the phenomenological result (\ref{tda}). The spectrum
for $D_\f$ cannot be predicted from the bare dimensional counting
and is justified within the framework of developed RG method.
We have pictured these spectrum  out in Fig. 4.

Being the solutions of RG-equations, the functions  $f_\f$ and $f_A$
are to be the arbitrary functions of the first integrals of (\ref{33}).
The number of first integrals is one less than  the number
 of arguments of $f$
 in (\ref{38}), and they can be
 founded from the system of equations
 \be
 \frac{ds}{s}=\frac{d\bar g}{\beta_g(\bar g,\bar u)}=
\frac{d\bar u}{\beta_u(\bar g,\bar u)},
 \label{39}
 \ee
supplied by some normalization
conditions for $\bar g$ and $\bar u.$ We use the standard one,
 \be
 \bar g(s=1,g,u)=g,\quad \bar u(s=1,g,u)=u.
  \label{40}
\ee
At the fixed point $N$ 3 from (\ref{39}) and (\ref{40})
one obtains the asymptotic solutions for $f_\f(s,g,u^2,z)$
 and $f_A(s,g,u^2, m^2r^{1/\nu})$:
\be
f_\f(s,g,u)= \left(\frac{s^{2\epsilon}\bar u^2}{u^2}\right)^{\eta_\f}
{\cal F}_\f(1,\bar u^2, \bar z),
 \label{41}
\ee
and
 \be
 f_\f(s,g,u)= \left(\frac{s^{2\epsilon}\bar u^2}{u^2}\right)^{\eta_A}
{\cal F}_A(1,\bar u^2,\bar z),
\label{42}
\ee
the scaling functions ${\cal F}_\f$ and ${\cal F}_A$ are not
fixed by the RG-equations and  calculated   usually
 in the framework of diagram technique.

\section{Short Distance Expansion. On the Possible
 Corrections to   Critical Spectra
  in the region $mr\to 0$}

Generally speaking, the existence of fixed points of RG-transformation
 does not  guarantee that the critical asymptotics (\ref{38}) do have place
in the real system.
As we have mentioned above, there would be another IR-divergences in the
scaling functions ${\cal F}_\f$ and ${\cal F}_A$
which are not handled by RG and, in principle,
 can modify the large-distance asymptotics close to the
region $m_0\ll k$.  By the way,
for $m_0/k\to 0,$ the critical dimensions  are not sufficient
 to derive  a conclusion on the long-range asymptotic behavior.
To investigate the model (\ref{A9}) with $m^2>0$ in the
  region $m_0\ll k$
 in details we use a Short Distance Expansion method following \cite{5}.

The short distance expansions of scaling functions
${\cal F}_\f$ and ${\cal F}_A$ in (\ref{41}) and (\ref{42}) provides us
 by an  asymptotic relation of the form
\be
{\cal F}(1, mr )=1+\sum_i
 c_i (r)m^{\Delta[O_i]},
\label{sde}
\ee
 where $\Delta [O_i]$ are the critical dimensions of all possible
statistical momenta
(the arbitrary products of fields and their derivatives averaged
with respect to one point)
 of various quantities of the theory (\ref{A9}).
In language of the modern critical phenomena theory such
 statistical momenta
is called as composite operators in analogy with the well-known
objects in quantum-field physics.

It is obvious that the  RG-predicted spectra (\ref{38})
are still  secure if
for all operators $\Delta[O_i]>0$.
The most important contributions into (\ref{sde}) for
 $mr\to 0$ are those of  the smallest $\Delta[O ]$.
In the framework of $\epsilon$-expansion
  $\Delta[O ]=d_O+{\cal O}(\epsilon),$
where $d_O$ is the canonical dimension of
 $O,$ therefore, if $\epsilon$ is small,
the canonical dimension $d_O$ provides
the major contribution to $\Delta[O ]$.
That is why, in principle, to justify the scaling laws
(\ref{38}), one can
limit the  checking of  critical dimensions
by the set of operators
with minimal canonical dimension $d_0.$
 At the  leading order we consider the
critical dimensions of the set of scalar quadratic
operators with  $d_O=1$:  $O_1=\f^2({\bf x})$
 and $O_2=A^2({\bf x}).$

The certain critical dimensions are assigned to
some linear combinations of
the operators $O_1$ and $O_2$ which
still invariant  in process of renormalization.
The basis of  renormalized composite operators
are related to that one of  non-renormalized operators
 through the  renormalization matrix
 $Z_{ik}$ such that  $F_i=Z_{ik}F^R_k$, \cite{8}.
In principle, the calculation of  matrix elements
requires the analysis of
diverging part of perturbation theory diagrams,
however, the use of gauge symmetry
consequences facilitates computations of the elements $Z_{ik}$
 substantially.
Since $O_1$ is a gauge invariant operator,
 but $O_2$ is not, the relevant Ward identities
 \cite{4} prove the triangle structure
for $Z_{ik}:$ $Z_{22}=1,$ $ Z_{21}=0.$ One can say
 that the
non-invariant operator $O_2$
does not contribute into scaling degrees
 of freedom  of the gauge invariant theory,
as well as it does not  admix to the gauge
 invariant operator $O_1$ in process of renormalization.

Furthermore, we need not compute diagrams to determine
 the element $Z_{11}.$
Acting by the differential operation
$\partial_{m^2}$ onto the partition functional
of renormalized theory (which is, obviously, finite
with respect to  the limit $\epsilon\to 0$),
we obtain the following  finite object
\be
\langle Z_3\f^2\rangle(\mb x),
\label{O1}
\ee
the finiteness of (\ref{O1}) leads to the relation
  $Z_{11}=Z^{-1}_3Z_m=Z_\f.$

In principle, (\ref{O1})
give us enough information to compute the complete set of
 critical dimensions of the
considered  statistical momenta. Since $Z_{ik}$ possesses a
triangle structure, exactly
the diagonal elements $Z_{kk}$ give the relevant anomalous correction:
\be
\gamma_{11}=\gamma_\f=\frac{\epsilon}{3},
\quad \gamma_{22}=0.
\label{ak}
\ee
 Both linear combinations of $O_1$ and $O_2$ which have the
definite scaling dimensions are also found unambiguously:
\be
\begin{array}{c}
C_1= O_1, \quad \Delta[O_1]=1+\frac {\epsilon}{3}, \\
C_2= O_1+a O_2,\quad \Delta[O_2]=d_O=1.
\end{array}
\label{lk}
\ee
The only reason that we need to compute
diagrams of perturbation theory for the element $Z_{12}$
is to determine the value of  $a$; this  calculation gives $a=2$
(this result is exact,{\em i.e.,} it still valid for any value
 of $\epsilon$)

For $\epsilon>0 $ the most important contribution to
(\ref{sde}) is provided by $C_2$: $\sim m$, and as it is  obvious,
this contribution  does not alter the scaling laws (\ref{38}).
The result (\ref{lk}) means that in the region  $mr\to 0$
  the scaling laws (\ref{38})   are still
secure as the universal characteristics of the theory.

 \section{  Spectrum of Energy}

The proposed model provides  a broad spectrum of
practical results which can be compared with  experimental data.
We now compute
the one-dimensional spectrum of energy of the fluid,
$E(r/l_p),$ versus the dimensionless distance apart
from the point of local perturbation.
As we have shown in the Introduction, the result on
this spectrum, in principle, can be derived from phenomenology,
(\ref{es}).
However, the question on justification
of the phenomenological result (\ref{es})
still remains, since the energy of   perturbed fluid
has two components,
\be
{E}(r/l_p)\equiv\frac 12
\langle(\partial\f)^2\rangle (r/l_p)+\frac 12
\langle F^2 \rangle (r/l_p),
\label{E}
\ee
which would   have dramatically different asymptotic
behavior in the large-distance limit. By the way,
(\ref{E}) would provide us an example of description
of a conserved integral quantity in terms of
scaling degrees of freedom.
The spectrum (\ref{E}) is governed by the statistical momenta
$E_\f\equiv\frac 12  \langle(\partial\f)^2\rangle (r/l_p) $
and $E_A\equiv\frac 12
\langle F^2 \rangle (r/l_p)$  of
the canonical dimension $d_{E}=d.$ To determine the
relevant critical indices we apply the trick which
we have employed in the previous Section:
Acting by the differential operations
 $m^2\partial_{m^2},$  $g\partial_g$, and $u\partial_u$
  onto the partition functional
of renormalized theory
 we obtain the following finite  objects
at the limit $\epsilon\to 0$:
\be
\left\{
\begin{array}{l}
\langle
 Z_3m^2\f^2
 \rangle( r/l_p), \\
\langle
 \left[
 g\partial_gZ_3
 \right] m^2\f^2+
\left
[g\partial_gZ_1
\right]
\left(E_\f+u^2\f^2A^2
\right)+
\frac 13
\left[
_4+g\partial_gZ_4
\right]g\f^4
\rangle (r/l_p),\\
\langle \left[u\partial_uZ_3\right] m^2\f^2+
\left[u\partial_uZ_1\right]E_\f+\frac 13\left[u\partial_uZ_4\right]
g\f^4 +  \frac 12 \left[u\partial_uZ_2\right] E_A + \\ +
 \left[ 2Z_1-u\partial_uZ_2\right]u^2\f^2A^2
\rangle (r/l_p).
\end{array}
\right.
\label{EE}
\ee
Any linear combination of (\ref{EE}) is again finite at the limit
$\epsilon\to 0,$ so that by means of simple
arithmetic operations one can derive from (\ref{EE})
the statements on finiteness for various
linear combinations of statistical momenta including
$E_\f  $ and $E_A.$

Taking into account that the scaling degrees of freedom
relevant to scalar wave modes coupling are vanished in the
large-distance limit ({\em i.e.,} assuming $g=0$ in (\ref{EE})),
 we obtain from (\ref{EE}) the
  combination containing $E_\f  $ and $E_A$:
\be
\langle
X_1E_\f+ \frac 12 X_2   E_A
\rangle (r/l_p),
\label{EEE}
\ee
where we have introduced $X_1\equiv\left[u\partial_uZ_1\right]$ and
  $X_2\equiv\frac 12\left[u\partial_uZ_2\right]$.   In the
framework of perturbation theory each of the
coefficients $X_i,$ as well as the  momenta
$E_\f$ and $E_A $
 have  poles in $\epsilon$, and consequently
 each term in (\ref{EEE}) separately  is  divergent
  if $\epsilon\to 0$.  The meaning of
 (\ref{EEE}) is that all the poles in $\epsilon$
   are subtracted out
 in such a way that the linear combination  in  (\ref{EEE})
  has a definite limit for $\epsilon\to 0$.

The linear form (\ref{EEE}) has two eigenvectors,
\be
 V_1=E_\f+ X_1^{-1}X_2E_A,\quad
 V_2=E_A+\frac 12 X_2^{-1}X_1 E_\f,
\label{IV}
 \ee
such that the poles in $\epsilon$ of $V_1$ are
subtracted out by those of $X_1$, and the poles
in $\epsilon$ of $V_2$ are  eliminated completely
by those in $X_2$. By the way, $\gamma_{V_1}=-D_\mu
\ln \left( u\partial_uZ_1\right)=-\epsilon/3$ and
$\gamma_{V_2}=-D_\mu
\ln \left( u\partial_uZ_2\right)=-2\epsilon.$

For the real value $\epsilon_r=1/2$ the major
contribution to the energy spectrum for the large-distance
asymptotics, $r/l_p\gg 1,$ is provided by
the combination from (\ref{IV}) of minimal critical dimension,
{\em i.e.,}
$V_2$  with
\be
\Delta [V_2]=
d+\gamma_{V_2}=d-1.
\label{result}
\ee
  Note, that $\Delta [V_1]=
d+\gamma_{V_1}=d-1/6.$ Furthermore, for $\epsilon_r=1/2$
$X_2^{-1}X_1=2$, and we have an explicit form
 for $V_1=E_\f+E_A.$
The last step of the computation is to perform
 a Fourier transformation
of  the momentum asymptotics with the index (\ref{result}) into
the real space $\{rl_p^{-1}\}$. Bringing it about,
 we, finally, obtain a
decaying profile (see Fig.5),
\be
{E}(r/l_p)\sim \left(\frac {l_p}{r}\right)^4
\label{end}
\ee
which meets the phenomenological result
presented in the Introduction.

\section{Conclusion}

The final conclusion is that in contrast with either
the statistical theory of waves (Zakharov) or
the theory of fully developed turbulence (Kolmogorov)
the  problem of coupling water to perturbation cannot
be solved from phenomenology in principle.
The matter is that the relevant  physical system
contains too many redundant degrees of freedom.
That is why to fix the statistically stable behavior
in the system one needs to  add
some extra assumptions on the character of perturbation
as well as on the character of wave modes coupling.

The problem of formulation of physically relevant
hypotheses on wave modes coupling mechanism can be
successfully solved by the use of various
quantum field theory techniques and
RG method in particular. This approach allows, first,
to integrate over the redundant physical degrees
of freedom, and, second, to investigate the
asymptotic properties of   physical systems
by means of analysis of their scaling degrees of freedom.

As a result we have formulated the effective action functional
which  allows to understand the water response for
a local external perturbation as a critical phenomenon.
The critical system of water coupled to perturbation
is subject to a "phase transition"
depending on the certain physical properties of
perturbation pulse.

The results on the asymptotic behavior derived from
the RG-analysis meet  those of partial results
which can be derived from  phenomenology.

\section{Acknowledgments}

One of the authors (D.V.) is grateful to
L. Volchenkova for fruitful discussions and checking
of particular computations.

\newpage

\vspace{6cm}

\begin{tabular}{|l|l|l|} \hline
\multicolumn{3}{|c|}{\bf Table 1.
  {\ } The fixed points of the model } \\ \hline
\bf N & \bf Coordinates
$\left\{ g'^*,u'^*\right\} $ & \bf The stability region
\\ \hline 1 & $\left\{ 0,0\right\}$ & $\epsilon<0$ \\
 \hline 2 & $\left\{ 0,\sqrt{\epsilon/3}\right\}$ &
unstable \\ \hline 3 & $\left\{ 0,-\sqrt {\epsilon/3}\right\}$ &
$0<\epsilon<3$ \\ \hline 4 &
$\left\{2\epsilon/3+(2\epsilon)^2\frac {17}{81},0\right\}$ &
 unstable \\ \hline 5 & $\left\{ -\frac{7}{34}
\left(1+\sqrt{1-\frac{816}{49}\left(\epsilon/3+
\sqrt{\epsilon/3}\right)}\right),
\sqrt {\epsilon/3}\right\}$ & unstable \\
\hline 6 & $\left\{-\frac{7}{34}\left(1+\sqrt{1-
\frac{816}{49}\left(\epsilon/3-\sqrt{\epsilon/3}\right)}\right),
-\sqrt {\epsilon/3}\right\}$ & unstable\\
\hline 7 & $\left\{ -\frac{7}{34}\left(1-
\sqrt{1-\frac{816}{49}\left(\epsilon/3+\sqrt{\epsilon/3}\right)}\right),
 \sqrt{\epsilon/3}\right\}$ & $0<\epsilon\leq 0.01$ \\
\hline 8 & $\left\{ -\frac{7}{34}\left(1-\sqrt{1-\frac{816}{49}
\left(\epsilon/3-\sqrt{\epsilon/3}\right)}\right),
 -\sqrt {\epsilon/3}\right\}$ & $3<\epsilon\leq 3.35$ \\
\hline
\end{tabular}

\newpage

CAPTIONS FOR FIGURES

FIGURE 1.

 The potential energy $U(f)$ versus $f$.a.)
For the case of $m^2>0$ the large-distance expectation
 value of $f$ is trivial;
 b.)  In the asymmetric case $m^2<0$ the classical minimum
 is degenerate.  There are two (in principle, an infinite
 number of) possible
expectation values for $f,$
$\pm \sqrt {\frac {|m^2|}{g}}$ Starting from a given minimum,
 it is possible to describe all other minima by
acting on the $f$ with the $U(1)$ symmetry group.

FIGURE 2.

  Symmetrical and asymmetric phases of the model with no coupling
 to vector potential $\mb A$;
a.)  $m^2>0$.  The net large-scale momentum of the
 fluid outside the eddy which is formed around the
perturbation region is equal to zero;
b.)  $m^2<0$.  The pressure pulse rises the net fluid
current into outside.  $P\ne 0.$

FIGURE 3.

 The diagram series of functions $\Sigma(p)$ in
the $f^4$-model in the critical phenomena theory.

FIGURE 4.

 The one-dimensional spectra for  the pair correlation functions
 of the theory   versus the dimensionless wave number.

FIGURE 5.

The one dimensional energy spectrum  $E(r/l_p)$ versus the
 dimensionless
distance apart from the perturbation point.

\end{document}